\documentclass[aps,prb,reprint,showpacs]{revtex4-2}
\usepackage{epsfig}\usepackage{subfigure}
\usepackage{dcolumn}
\usepackage{graphicx,graphics}
\usepackage{amsmath,ulem}
\usepackage{amssymb}
\usepackage{bm}
\usepackage{color}
\usepackage[utf8]{inputenc}
\usepackage{pifont}
\usepackage[colorlinks,linkcolor={blue},citecolor={blue},urlcolor={blue}]{hyperref}
\usepackage{mathtools}
\usepackage{mathtools}
\usepackage{bigints}
\newcommand{\bk}{{\bm k}}

\newcommand{\kp}{k_{\tiny\parallel}}
\newcommand{\kpb}{{\bm k}_{\tiny\parallel}}

\begin{document}

\author{Mahmoud M. Asmar}
\affiliation{Department of Physics and Astronomy, The University of Alabama, Tuscaloosa, AL 35487, USA }
\author{Gaurav Gupta}
\affiliation{Department of Physics and Astronomy, The University of Alabama, Tuscaloosa, AL 35487, USA }
\author{Wang-Kong Tse }
\affiliation{Department of Physics and Astronomy, The University of Alabama, Tuscaloosa, AL 35487, USA }

\title{ Particle-hole asymmetry and quantum confinement effects on the magneto-optical response of topological insulator thin-films   }
\date\today

\begin{abstract}
Intrinsically broken symmetries in the bulk of topological insulators (TIs) are manifested in their surface states. In spite of particle-hole asymmetry in TIs, it has often been assumed that their surface states are characterized by a particle-hole symmetric Dirac energy dispersion. In this work we demonstrate that the effect of particle-hole asymmetry is essential to correctly describe the energy spectrum and the magneto-optical response in TIs thin-films.
In thin-films of TIs with a substantial degree of particle-hole symmetry breaking, such as Sb$_2$Te$_3$, the longitudinal optical conductivity displays absorption peaks arising from optical transitions between bulk and surface Landau levels for low photon energies. The transition energies between the bulk and surface Landau levels exhibit clearly discernable signatures from those between surface Landau levels due to their distinct magnetic field dependence. Bulk contributions to the magneto-optical conductivity in a TI thin-film are enhanced via one type of doping while being suppressed by the other. This asymmetric dependence on type of doping aids in revealing the particle-hole asymmetry in TI thin-films.
\end{abstract}

\maketitle

\section{Introduction}
Three-dimensional topological insulators (TIs) are bulk insulating materials that host topologically protected semimetallic states at their surfaces~\cite{Rev1,Rev2,BHZmodel,Zhang2009,Liu,ARPES1,ARPES3,ARPES2,ARPES5,ARPES4,ARPES6}. A Dirac-like quasiparticle dispersion characterizes the surface states of TIs, and a two-dimensional Dirac-like Hamiltonian effectively describes these low energy excitations. The existence of such topologically protected, Dirac-like, surface states was first established in angle-resolved photo emission spectroscopy (ARPES) experiments on materials such as Bi$_2$Se$_3$, Bi$_2$Te$_3$ and Sb$_2$Te$_3$~\cite{ARPES1,ARPES3,ARPES2,ARPES5,ARPES4,ARPES6}.

The ability to manipulate and control the properties of TIs surface states is crucial to incorporating these materials in versatile applications such as spintronics and quantum computing~\cite{sinova,spintronics,macdonald,Burkov1}.  The existence of surface states in three-dimensional TIs is guaranteed by topology. In contrast, the properties of the fermions residing at the surface are influenced by intrinsically broken symmetries in the TI bulk, reduced surface symmetries, and by the TI thickness ~\cite{mahdavan1,mahdavan2,interfacesymmery,dav,thingap3,thingap2,thingap1,thingap0,central}. Tuning the topological surface states properties via thickness variations of TI thin-films is a promising route made possible by the advancements in synthesizing ultra thin-films of these exotic materials ~\cite{yuxuan,nanorib,assymetry,bulk2}. Finite-size effects lead to a surface spectral gap due to the hybridization of the surface states residing at opposite ends via their bulk extensions~\cite{thinn0,thinn1,thinn2,thingap3,thingap2,thingap1,thingap0,central}. Theories describing the electronic properties of TI thin-films are based on effective surface descriptions that often neglect the TI's intrinsically broken particle-hole (p-h) symmetry~\cite{thingap2,thingap1,thingap0,thinB,magopt1,magopt2} and quantum confinement~\cite{thingap3}, both of which are essential in the interpretation of experiments probing the electronic properties of TI thin-films ~\cite{yuxuan}.

A powerful method for probing electron dynamics is strong-field magneto-optical spectroscopy.  The measurement of the magneto-optical conductivity provides an accurate characterization of the Landau level structures ~\cite{graphite1,graphite2,dcond2,dcond3}. Motivated by the tunable electronic properties of TI thin-films and the precise electronic characterization provided by magneto-optical spectroscopy, we theoretically study and contrast the magneto-optical conductivity of Sb$_2$Te$_3$ and Bi$_2$Se$_3$ TI thin-films. In this work, we start by characterizing the main properties of the TI's surface states that have a direct impact on the TI thin-film energy spectrum. These properties include the dependence of the surface states' electronic dispersion and their extension into the bulk on the intrinsically broken p-h symmetry. Self-consistently considering the TI bulk Hamiltonian and the termination imposed boundary conditions in a thin-film, we find the effects of quantum confinement and p-h asymmetry on the surface and bulk states of Sb$_2$Te$_3$ and Bi$_2$Se$_3$ thin-films. Considering a quantizing magnetic field perpendicular to the TI thin-film termination, we determine the bulk and surface Landau levels spectrum. We contrast the salient differences of the Sb$_2$Te$_3$ and Bi$_2$Se$_3$ thin-films Landau levels. These differences reflect the degrees of p-h asymmetry in TI thin-films, and impact their magneto-optical response. Our analysis of the Landau levels optical transitions energies and the absorptive part of the longitudinal conductivity shows that TI thin-films with strong p-h asymmetry (Sb$_2$Te$_3$), display low-energy absorption peaks originating from bulk-to-surface Landau levels transitions. Conversely, bulk-to-surface Landau levels optical transitions are not manifested in systems with a low degree of p-h symmetry breaking (Bi$_2$Se$_3$) as their absorption peaks are absent in the dynamic longitudinal conductivity for low photon energies. Analyzing the transition energies dependence on magnetic field strength, we show that the bulk-to-surface optical transitions do not display the typical dependence expected from Dirac Fermions. We also suggest that the optical conductivity absorption peaks arising from bulk-to-surface transitions facilitate the determination of the shifted location of the Dirac point in p-h asymmetric systems.

The remainder of the paper is devoted to the effects of p-h symmetry breaking and quantum confinement in thin-films of TIs and the manifestation of these effects in their magneto-optical response. In Sec.~\ref{sec1} we first analyze the influence of p-h asymmetry on the surface states of TIs and their extensions into the bulk. Then we introduce a self-consistent method to determine the bulk and surface states in a TI thin-film, which allows for an analysis of quantum confinement effects in this system. We consider a quantizing magnetic field in Sec.~\ref{sec2} and self-consistently determine the surface and bulk Landau levels for the thin-film. In Sec.~\ref{sec3} we calculate the absorption component of the optical conductivity of TIs thin-films, and show the effects of the intrinsic p-h asymmetry on the Landau levels transition energies of the absorption peaks of the longitudinal optical conductivity. We present our summary, conclusions, and final remarks in Sec.~\ref{sec4}.

\section{Thin-films of Topological Insulators }\label{sec1}
In this section we introduce and review the properties of TI surface states in semi-infinite TI samples and TI thin-films, and demonstrate the crucial role of p-h symmetry breaking in these systems~\cite{thingap0,thingap1,thingap2,thingap3,Liu}. We start by recalling that
the low energy bulk excitations of three-dimensional TIs, Bi$_2$Se$_3$, Bi$_2$Te$_3$ and Sb$_2$Te$_3$, are centered in the vicinity of the $\Gamma$ point of the Brillouin zone. The electron dynamics of these bulk excitations are captured by the effective $\bk \cdot {\bm p}$ Hamiltonian~\cite{Liu}. In the parity $(+,-)$ and spin $(\uparrow,\downarrow)$ basis, $\psi^{T}=(\psi_{+\uparrow},\psi_{-\uparrow},\psi_{+\downarrow},\psi_{-\downarrow})$, this Hamiltonian reads
\begin{eqnarray}\label{BTI}
&&H_{\rm TI}(\bk)=\\
&&\sigma_{0}\tau_{0}\epsilon(\bk)+ \sigma_{0}\tau_{z}\mathcal{M}(\bk)+A_{1}\sigma_{0}\tau_{y}k_{z}+A_{2}({\bm \sigma}\times \mathbf{k})_{\hat{z}}\tau_{x}.\nonumber
\end{eqnarray}
Here $\epsilon(\bk)=C+D_{1}k^{2}_{z}+D_{2}\kp^{2}$, $\mathcal{M}(\bk)=M-B_{1}k_{z}^2-B_{2}\kp^{2}$, $\kp^{2}=k^{2}_{x}+k^{2}_{y}$, the Pauli matrices $\sigma$ ($\tau$) act in the spin (parity) space, with $\tau_{0}$ and $\sigma_{0}$ the identity matrices, and the parameter set $(A_{i},B_{i},D_{i},M,C)$ is material dependent.
While the Hamiltonian in Eq.~\eqref{BTI} is time-reversal invariant, $[\mathcal{T},H_{\rm TI}(\bk)]=0$ where $\mathcal{T}=i\tau_{0}\sigma_{y}\mathcal{C}$ and $\mathcal{C}$ is the antilinear conjugation operator, it does not possess particle-hole (p-h) symmetry since $\{\mathcal{P},H_{\rm TI}(\bk)\}\ne 0$, where $\mathcal{P}=\tau_{x}\sigma_{z}$~\cite{Ryu,Chamon}. The lack of p-h symmetry in  $H_{\rm TI}$ is directly attributed to the non-zero p-h asymmetric terms which are captured by $\epsilon(\bk)$ in Eq.~\eqref{BTI}.

The termination of a TI leads to the appearance of topologically protected surface states. For a TI thin-film, the states residing at opposite surfaces hybridize due to their extension into the bulk, leading to a gap in the surface states spectrum. To gain a clear picture of the TI material parameters governing the thin-film electronic properties, it is instructive to start by analyzing the energy dispersion and $z$-extension of the surface states of TIs in their semi-infinite limit.
When the extension of the TI along its termination direction is large enough so that finite-size effects can be neglected, the TI can be considered semi-infinite, {\it i.e.}, terminated along its [111]-crystallographic direction at $z=0$ and extending to $z=-\infty$. Translational invariance in the $x-y$ plane leads to a conserved in-plane momentum, $\kpb$, while in the $z$-direction $k_{z}=-i\partial_{z}$. Moreover, the TI Hamiltonian in Eq.~\eqref{BTI} preserves helicity, such that $[\hat{h},H_{\rm TI}(\bk)]=0$, where $\hat{h}=({\bm \sigma}\times \mathbf{k})_{\hat{z}}/\kp$ is the helicity operator, and the bulk states of $H_{\rm TI}$ can be categorized according to their helicity, $\hat{h}\psi_{t}(\kpb,z)=t\psi_{t}(\kpb,z)$ with $t=\pm1$. Helicity conservation and the boundary condition $\psi_{t}(\kpb,z\rightarrow-\infty)=0$ aid in the construction of the helicity ansatz of the TI states, $\psi^{T}_{t,\mu}(\kpb,z)=\Phi^{T}_{t,\mu}e^{i\kpb\cdot{\bm r }}e^{\lambda_{\mu} z}$, where
\begin{equation}\label{phi}
\Phi^{T}_{t,\mu}=  (ia_{t,\mu},ib_{t,\mu},ta_{t,\mu}e^{i\theta_{k}},tb_{t,\mu}e^{i\theta_{k}})^{T},
\end{equation}
$\tan(\theta_{k})=k_y/k_x$, and the coefficients $a_{t,\mu}$ and $b_{t,\mu}$ are dependent on $\lambda_{\mu}$. Applying the helical ansatz, $\psi_{t,\mu}(\kpb,z)$, to the Hamiltonian, $H_{\rm TI}(\kpb,-i\partial_{z})\psi_{t,\mu}(\kpb,z)=E\psi_{t,\mu}(\kpb,z)$, reduces the problem to
\begin{eqnarray}\label{eign1}
[\tau_{0}\epsilon(\kpb,i\lambda_{\mu})+\tau_{z}\mathcal{M}(\kpb,i\lambda_{\mu})+tA_{2}\tau_{x}\kp &&+iA_{1}\tau_{y}\lambda_{\mu}]\bar{\psi}_{t,\mu}\nonumber\\&&=E\bar{\psi}_{t,\mu}\;,
\end{eqnarray}
where $\bar{\psi}^{T}_{t,\mu}=(a_{t,\mu},b_{t,\mu})^{T}$.
It follows from Eq.~\eqref{eign1} that the decay parameter, $\lambda_{\mu}(E)$, is energy dependent as it satisfies $[\epsilon(\kpb,i\lambda_{\mu})-E]^2-\mathcal{M}^{2}(\kpb,i\lambda_{\mu})=A^{2}_{1}\lambda^{2}_{\mu}-A^{2}_{2}\kp^{2}$, and the coefficients $a_{t,\mu}= \mathcal{M}(\kpb,i\lambda_{\mu})-\epsilon(\kpb,i\lambda_{\mu})+E$ and $b_{t,\mu}=A_{1}\lambda_{\mu}+tA_{2}\kp$. Solving the characteristic equation for $\lambda_{\mu}(E)$ results in four distinct values of these decay parameters. The boundary condition $\psi_{t}(\kpb,z\rightarrow-\infty)=0$ restricts the $\lambda_{\mu}(E)$ solutions to two values satisfying ${\rm Re}[\lambda_{\mu}(E)]>0$, and we differentiate them by the index $\mu=\pm1$. Additionally, conservation of helicity allows solving the boundary condition at $z=0$, $\sum_{\mu}C_{\mu}\psi_{t,\mu}(\kpb,0)=0$, for each helicity. The states satisfying the $z=0$ boundary condition are the TIs surface states and their energies are constrained to
\begin{eqnarray}\label{sdisp}
E_{t}(\kp)=C+\frac{D_{1}M}{B_{1}}+&&tA_{2}\sqrt{1-\left(\frac{D_{1}}{B_{1}}\right)^{2}}\kp\nonumber\\
&&+\left(D_{2}-\frac{D_{1}B_{2}}{B_{1}}\right)\kp^{2}\;,
\end{eqnarray}
The equation above indicates that the surface states' Dirac point is located at $E_{\rm D}=C+D_{1}M/B_{1}$, the Fermi velocity $ v_{{\rm F}}=(A_{2}/\hbar) \sqrt{1-(D_{1}/B_{1})^2}$, and the effective mass is $\hbar^{-2} d^{2}E_{s}/d\kp^2=m^{-1}_{s}=[D_{2}-D_{1}B_{2}/B_{1}]/\hbar^2$.
\begin{figure}[t]
  \centering
  \includegraphics[width=\columnwidth]{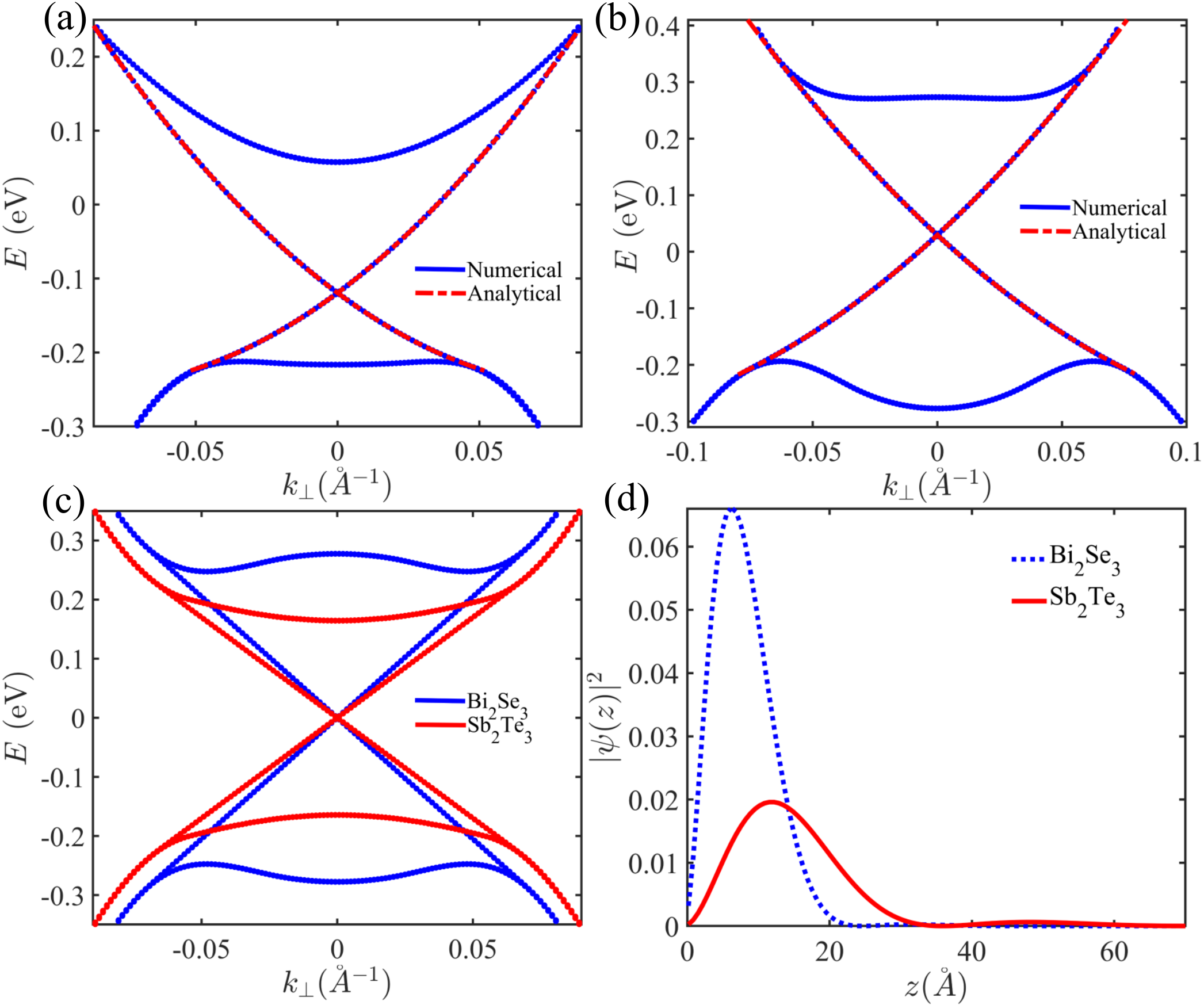}
  \caption{Surface and bulk bands near the $\Gamma$ point for a TI semi-infinite sample of (a) Sb$_2$Te$_3$ and (b) Bi$_2$Se$_3$. (c) Bulk and surface dispersions neglecting p-h asymmetry in  Sb$_2$Te$_3$  and Bi$_2$Se$_3$. (d) Bulk extension of the surface states of the TIs in (a) and (b) with a Fermi energy at the Dirac point.}\label{fig1}
\end{figure}
Hence, it becomes apparent that the dispersion relation of the surface states is strongly dependent on the p-h asymmetric bulk parameters leading to clear differences between different TIs. For example, in Bi$_2$Se$_3$ $E_{{\rm D},{\rm Bi_2Se_3}}=0.03$ eV, $ v_{{\rm F},{\rm Bi_2Se_3}}=6.2\times10^{5}$ m/s, $m_{{\rm Bi_2Se_3}}=0.68m_{0}$, where $m_{0}$ is the electron's rest mass. On the other hand, for  Sb$_2$Te$_3$ the Dirac point is located at $E_{{\rm D},{\rm Sb_2Te_3}}=-0.119$ eV, $v_{{\rm F},{\rm Sb_2Te_3}}=4.3\times 10^{5}$ m/s, and $m_{{\rm Sb_2Te_3}}=0.53m_{0}$. Here we have adopted the Bi$_2$Se$_3$ and Sb$_2$Te$_3$ $\bk\cdot {\bm p}$ parameters extracted from first-principle calculations and experimental fits~\cite{Zhang2009,Liu,yuxuan}. In Ref.~\cite{note} we provide the numerical values of the $\bk\cdot {\bm p}$ parameters for Bi$_2$Se$_3$ and Sb$_2$Te$_3$. From Fig.~\ref{fig1}(a-b) and Eq.~\eqref{sdisp} we notice that the p-h asymmetric terms in the TI Hamiltonian, Eq.~\eqref{BTI}, not only shift the Dirac point's location from $E=0$ but also introduce curvature to the states’ dispersion while reducing their Fermi velocity. Neglecting these p-h asymmetric terms leads to fictitious similarities between the states corresponding to different materials, as shown in Fig.~\ref{fig1}(c). Since the TI surface states extend into the bulk of the TI, to understand this extension we analyze the decay parameters evaluated at the Dirac point~\cite{Liu}. $\lambda_{\mu}(E_{\rm D})$ for Bi$_2$Se$_3$ and Sb$_2$Te$_3$, where $E_{\rm D}\approx D_{1}M/B_{1}$, takes the form
\begin{equation}\label{decay}
\lambda_{\mu}(E_{\rm D})=\frac{1}{2|B_{1}|}\left[\frac{A_{1}}{\sqrt{\mathcal{R}}}+\mu \sqrt{\frac{A^{2}_{1}}{\mathcal{R}}-4B_{1}M}\right].
\end{equation}
Here
\begin{equation}\label{R}
\mathcal{R}=1-\left(D_{1}/B_{1}\right)^2\;,
\end{equation}
captures the p-h symmetry breaking effects. The form of $\lambda_{\mu}(E_{\rm D})$ and the boundary condition $\psi_{t}(z\rightarrow-\infty)=0$, capture the conditions on the material parameters that lead to surface states. For a surface state to exist we need ${\rm Re}[\lambda_{\mu}(E_{\rm D})]>0$, {\it i.e.}, the states are normalizable. From Eq.~\eqref{decay} and assuming $A_{1}>0$, which is satisfied for both Bi$_2$Se$_3$ and Sb$_2$Te$_3$, we can notice that the surface state wavefunctions are normalizable if $B^{2}_{1}-D^{2}_{1}>0$ and $MB_{1}>0$. Moreover, if the latter conditions are satisfied and $A^2/\mathcal{R}<4B_{1}M$, $\lambda_{\mu=\pm}(E_{D})$ are complex conjugate partners, which is the case in Bi$_2$Se$_3$ and Sb$_2$Te$_3$~\cite{note}.
In general, the presence of two decay parameters result in normalizable states and makes the boundary condition, $\sum_{\mu}C_{\mu}\psi_{t,\mu}(z=0)=0$, non-trivially satisfied. The states satisfying the boundary condition are helical surface states given by $\psi_{s}(E_{t})=\mathcal{A}(e^{\lambda_{+}z} -e^{\lambda_{-}z})\Phi_{t}$, where $\mathcal{A}$ is a constant, $\Phi_{t}=\Phi_{t,+}=\Phi_{t,-}$, $\Phi_{t,\mu}$, Eq.~\eqref{phi}, since $a_{t,+}=a_{t,-}$ and $b_{t,+}=b_{t,-}$ for the surface states energy $E_{t}$, Eqs.~\eqref{eign1}-\eqref{sdisp}. In addition to their helical nature the surface states decay and oscillate into the bulk of the TI because $\lambda_{\mu}(E_{D})$ is complex, Fig.~\ref{fig1}(d). The oscillations wavelength of the surface states into the bulk direction is $\Upsilon=2\pi/ {\rm Im}[\lambda_{+}(E_{D})]$, such that
\begin{equation}\label{period}
  \Upsilon=4 \pi |B_{1}|{\bigg /}\sqrt{\left|\frac{A^{2}_{1}}{\mathcal{R}}-4B_{1}M\right|}\;.
\end{equation}
The $z$-extension of the surface states is governed by their decay length, $l_{d}$, which is given by
$l_{d}={\rm Re}[1/\lambda_{\pm}(E_{D})]$, {\it i.e.},
\begin{equation}\label{dlength}
l_{d}=2|B_{1}|\sqrt{\mathcal{R}}/A_{1}\;.
\end{equation}
For Bi$_2$Se$_3$ the surface states decay length is $l_{d}=9.1$ \AA, while $l_{d}=21$ \AA\; for Sb$_2$Te$_3$. The difference in the decay lengths of Bi$_2$Se$_3$ and Sb$_2$Te$_3$ is reflected in the localization and extension of their states into the bulk. As shown in Fig.~\ref{fig1}(d), the surface states of Bi$_2$Se$_3$ are maximally localized within the first quintuple layer (QL) and extend up to the second QL~\cite{Karunya}. On the other hand, the Sb$_2$Te$_3$ surface states maximally localize within the second QL and have non-vanishing weights up to 5QLs. We also notice that p-h asymmetry leads to shorter decay length, $l_{d}$, since $0<\mathcal{R}<1$.

The consideration of realistic material parameters in the $\bk\cdot {\bm p}$ Hamiltonian of Sb$_2$Te$_3$ and Bi$_2$Se$_3$, such as the p-h asymmetric terms, leads to striking differences between their surface states' energy-momentum dispersion and the extension of these states into the bulk. These bulk-inherited differences become even more relevant when studying a TI thin-film. Consider a free-standing TI thin-film of thickness $d=2a$ and terminated along its $z$-direction at $z=a$ and $z=-a$. The TI thin-film is described by the Hamiltonian,
\begin{equation}\label{Hthin}
  H_{{\rm TF}}(\kpb,z)=H_{\rm TI}(\kpb,z)\Theta(z+a)\Theta(a-z)\,
\end{equation}
where $\Theta$ is the Heaviside step function. The TI thin-film lacks translational invariance along the $z$-direction and $k_{z}=-i\partial_{z}$, but it preserves both the in-plane momentum $\kpb$ and the $z$ to $-z$ inversion, since $\mathcal{I} H_{{\rm TF}}(\kpb,z)\mathcal{I} = H_{{\rm TF}}(\kpb,-z)$ where the unitary operator $\mathcal{I}=\tau_{z}\sigma_{z}$.

\begin{figure}[t]
  \centering
  \includegraphics[width=\columnwidth]{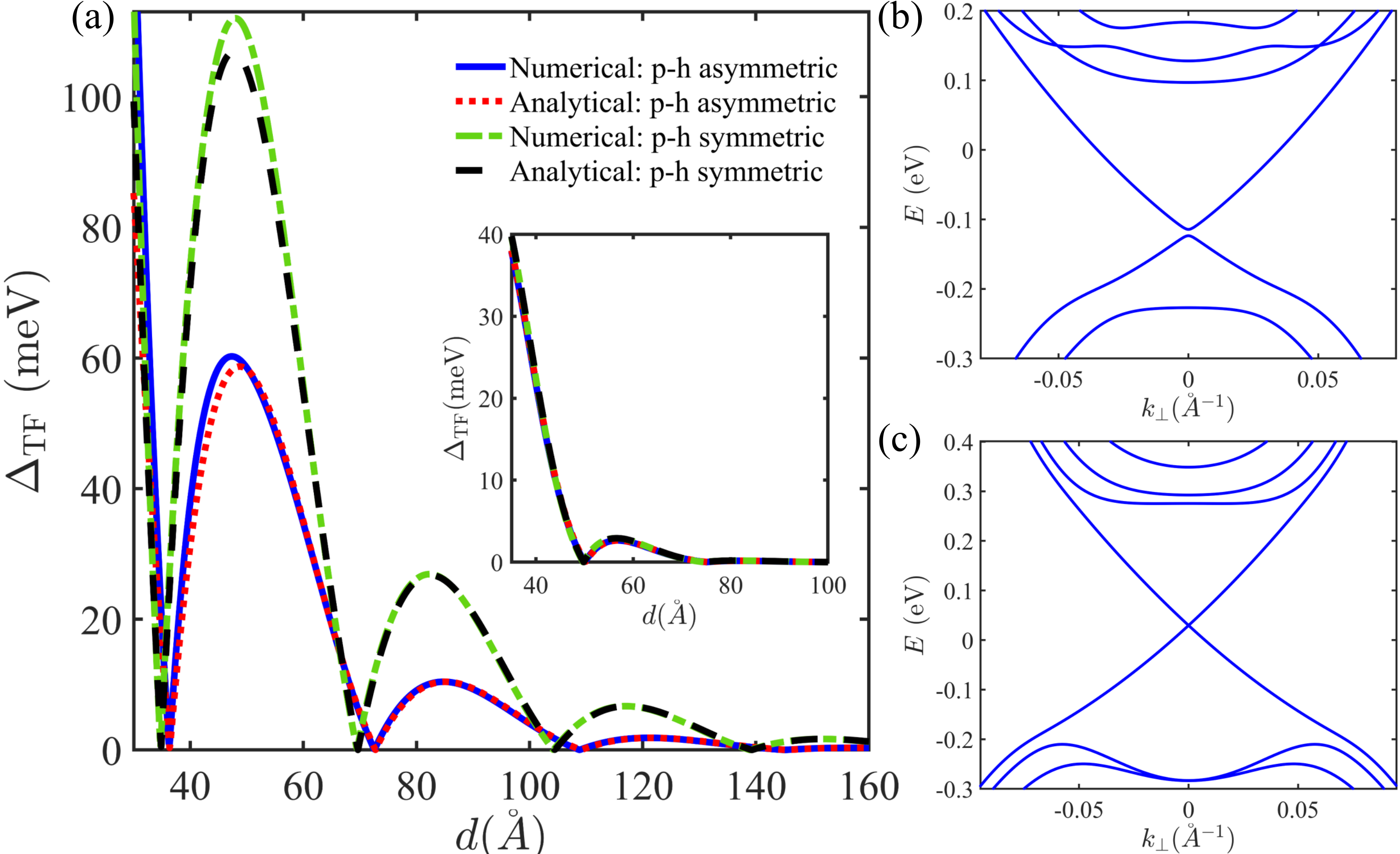}
  \caption{(a) Thickness dependence of the hybridization gap, $\Delta_{\rm TF}$, generated in a Sb$_2$Te$_3$ thin-film, with and without p-h asymmetry. The inset of (a) shows  $\Delta_{\rm TF}$ as a function of film thickness for Bi$_2$Se$_3$. Panels (b) and (c) show the low-energy dispersion relation of an $8$ QL thin-film of Sb$_2$Te$_3$ and Bi$_2$Se$_3$.}\label{fig2}
\end{figure}

To discern p-h symmetry breaking and finite-size effects on the properties of the surface and bulk states of the TI thin-film, we need to self-consistently solve for the Hamiltonian's bulk states, and the termination imposed boundary condition, $\psi(\kpb,a)=0$ and $\psi(\kpb,-a)=0$. Numerically, it is convenient to adopt the central-finite difference method to obtain the bulk and surface eigenvalues and eigenvectors of the TI thin-film~\cite{central}. In this method the terminated direction is discretized, for the thin-film in Eq.~\eqref{Hthin} that is the $z$-direction, such that $z_{n}=n h_z$, where $h_{z}$ is the discretization step, $n=0,1,\hdots N-1$, and $Nh_z=2a$. Similarly, the spinors, written in the basis of Eq.~\eqref{BTI}, take discrete values $\psi_{n}(\kpb)=\psi(\kpb,z_{n})$ and the differential operators are replaced by central finite differences, such that $\partial_{z}\psi_{n}(\kpb)=[\psi_{n+1}(\kpb)-\psi_{n-1}(\kpb)]/(2h_{z})$, $\partial^{2}_{z}\psi_{n}(\kpb)=[\psi_{n+1}(\kpb)+\psi_{n-1}(\kpb)-2\psi_{n}(\kpb)]/h^{2}_{z}$. The boundary condition for a vacuum terminated TI thin-film is imposed by setting $\psi_{-1}(\kpb)=\psi_{N}(\kpb)=0$. Hence, performing the discrete mapping of Eq.~\eqref{Hthin} reduces the boundary value problem to finding the eigenvalues and vectors of a block-tridiagonal, tight-binding-like, matrix
\begin{eqnarray}\label{discH}
&&\left[H_{{\rm TF}}(\kpb)\right]_{m,n}=\\
&&\sum_{\alpha=\pm1}\sigma_{0}\left[-\tau_{0}\frac{D_{1}}{h^{2}_{z}}+\tau_{z}\frac{B_{1}}{h^{2}_{z}} -i\alpha\tau_{y}\frac{A_{1}}{2h_{z}}\right]\delta_{m+\alpha,n}\nonumber\\
&&+\left[\sigma_{0}\tau_{0}\epsilon_{\perp}+\sigma_{0}\tau_{z}\mathcal{M}_{\perp}+ A_{2}({\bm \sigma}\times \mathbf{k})_{\hat{z}}\tau_{x}\right]\delta_{m,n}\nonumber\;.
\end{eqnarray}
We have defined $\epsilon_{\perp}$$=$$\epsilon(\kpb,k_{z}=0)-2D_{1}/h^{2}_{z}$ and $\mathcal{M}_{\perp}=\mathcal{M}(\kpb,k_{z}=0)+2B_{1}/h^{2}_{z}$. Diagonalizing the discretized boundary value problem results in the set of eigenvalues and eigenvectors that satisfy both the Hamiltonian and the boundary conditions in Eq.~\eqref{Hthin}. The numerically evaluated surface states' gap in the thin-film, $\Delta_{\rm TF}$, is shown in Fig.~\ref{fig2}(a)
for Sb$_2$Te$_3$ and Bi$_2$Se$_3$(inset). In these two materials $\Delta_{\rm TF}(d)$ decays and oscillates with thickness $d$, however, in Sb$_2$Te$_3$ this gap is significantly larger than in Bi$_2$Se$_3$ for the same range of $d$. When $d$ is larger than the decay length, $l_{d}$, Eq,~\eqref{dlength}, we can find an approximate analytical form for the spectral gap of the surface states and discern the crucial material parameters influencing this gap. Finding $\Delta_{\rm TF}$ in the regime $d/l_{d}\gg 1$ can be done by adopting the tunneling approximation which assumes that the states residing at opposite surfaces of the TI thin-film weakly hybridize via their $z$-extensions and generate a surface spectral gap~\cite{thingap0,thingap2}. The surface states gap for $d/l_{d} \gg1$ takes the form
\begin{equation}\label{agap}
\Delta_{{\rm TF}}(d)=\frac{4\Upsilon\mathcal{R}|M| }{\pi l_{d}}\left|\sin\left(2\pi d\right/ \Upsilon )\right|e^{-d/l_{d}}\;,
\end{equation}
where $\mathcal{R}$, $\Upsilon$, and $l_{d}$ are given in Eqs.~\eqref{R}, \eqref{period} and \eqref{dlength}. It is interesting to notice that similar to the $z$-dependence of the surface states, Eqs.~\eqref{decay}-\eqref{dlength}, the thin-film gap also oscillates in its terminated direction, $z$, with a wavelength $\Upsilon$ and has a decay length of $l_{d}$. In Fig.~\ref{fig2}(a), we notice conspicuous p-h symmetry breaking effects in the gap of the Sb$_2$Te$_3$ while these effects are almost negligible for the Bi$_2$Se$_3$ gap. The latter is attributed to the strong difference between $\mathcal{R}$ for these two materials. In Bi$_2$Se$_3$, $\mathcal{R}\approx 1$ which means that the decay parameter, $\lambda_{\mu}(E_{\rm D})$ in Eq.~\eqref{decay}, is approximately that of the p-h symmetric case. Since $\lambda_{\mu}(E_{\rm D})$ determines both $\Upsilon$ and $\l_{d}$, then, for  Bi$_2$Se$_3$ these two parameters are approximately equal in the p-h symmetric and antisymmetric cases and, consequently, the effects of the p-h asymmetric terms are negligible in $\Delta_{\rm TF}$ for this material, as shown in the inset of Fig.~\ref{fig2}(a). In Sb$_2$Te$_3$ the situation is very different since $\mathcal{R}<1$. This leads to a smaller spectral gap oscillations amplitude $4\Upsilon\mathcal{R}|M|/(\pi l_{d})$, and a shorter decay lengths $l_{d}$  due to p-h asymmetry. The latter two effects make $\Delta_{{\rm TF}}(d)$ in the p-h asymmetric case smaller than that of the p-h symmetric case, as shown in Fig.~\ref{fig2}(a). In addition to the gap generated in the surface states spectrum the thin-film geometry leads to the bulk states quantization due to quantum confinement. This effect is not captured in effective surface models as it requires a self-consistent treatment of the bulk states and the boundary conditions, {\it i.e}, Eq.~\eqref{discH}. In Figs.~\ref{fig2}(b) and (c), we show the surface and bulk spectrum for an 8QL thin-film of Sb$_2$Te$_3$ and Bi$_2$Se$_3$ obtained from our self-consistent calculation, Eq.~\eqref{discH}. In Fig.~\ref{fig2}(b) we notice the surface gap of Sb$_2$Te$_3$, and a clear quantization of its bulk states due to quantum confinement. For Bi$_2$Se$_3$, the surface gap is almost negligible, still, the quantum confinement of its bulk states is clearly non-negligible, Fig.~\ref{fig2}(c).

This section has addressed the influence of p-h symmetry breaking and quantum confinement effects on the surface and bulk states of TIs. We noticed that the p-h symmetry breaking effects are more pronounced in Sb$_2$Te$_3$ than in Bi$_2$Se$_3$. We have also shown that in order to accurately capture the states and energy spectrum governing TI thin-films' electronic properties, a self-consistent approach to the thin-film bulk and boundary should be adopted. In what follows we adopt the self-consistent method developed in this section to study the effects of p-h symmetry breaking and quantum confinement on the Landau levels quantization and magneto-optical response of TI thin-films.

\section{External Magnetic Field}\label{sec2}
The effects of an external magnetic field, $B$, perpendicular to the TI thin-film are captured by two additional terms in the thin-film Hamiltonian, Eq.~\eqref{Hthin}. The first term describes the orbital coupling to magnetic field through the minimal coupling $k_{x}\rightarrow k_{x}+(e/\hbar)A_{x}$, where ${\bm A}=[-By,0]$ is the vector potential in the Landau gauge. The second term describes the coupling of the spins to the magnetic field and is given by the Zeeman contribution $H_{\rm Z}(B)=(\mu_{\rm B}B/2)\left(g_{-}\sigma_{z}\tau_{z}+g_{+}\sigma_{z}\tau_{0}\right)$, where $g_{\pm}=(g_{1}\pm g_{2})/2$, $g_{1,2}$ are the effective g-factors, and $\mu_{\rm B}$ is the Bohr magneton~\cite{note}. The Hamiltonian in Eq.~\eqref{Hthin} then becomes,
\begin{eqnarray}\label{HThinB}
&&H_{{\rm TF,B}}(\kpb,z)=\\
&&\left[H_{\rm TI}(k_{x}-eBy/\hbar,k_{y},z)+H_{\rm Z}(B)\right]\Theta(z+a)\Theta(a-z)\nonumber\;.
\end{eqnarray}
We define the lowering and raising operators $a=l_{\rm B}[\partial_{y}+(y-y_{0})/l^2_{\rm B}]/\sqrt{2}$ and $a^{\dag}=l_{\rm B}[-\partial_{y}+(y-y_{0})/l^2_{\rm B}]/\sqrt{2}$ with $l_{\rm B}=\sqrt{\hbar /(e|B|)}$
\begin{figure}[t]
  \centering
  \includegraphics[width=\columnwidth]{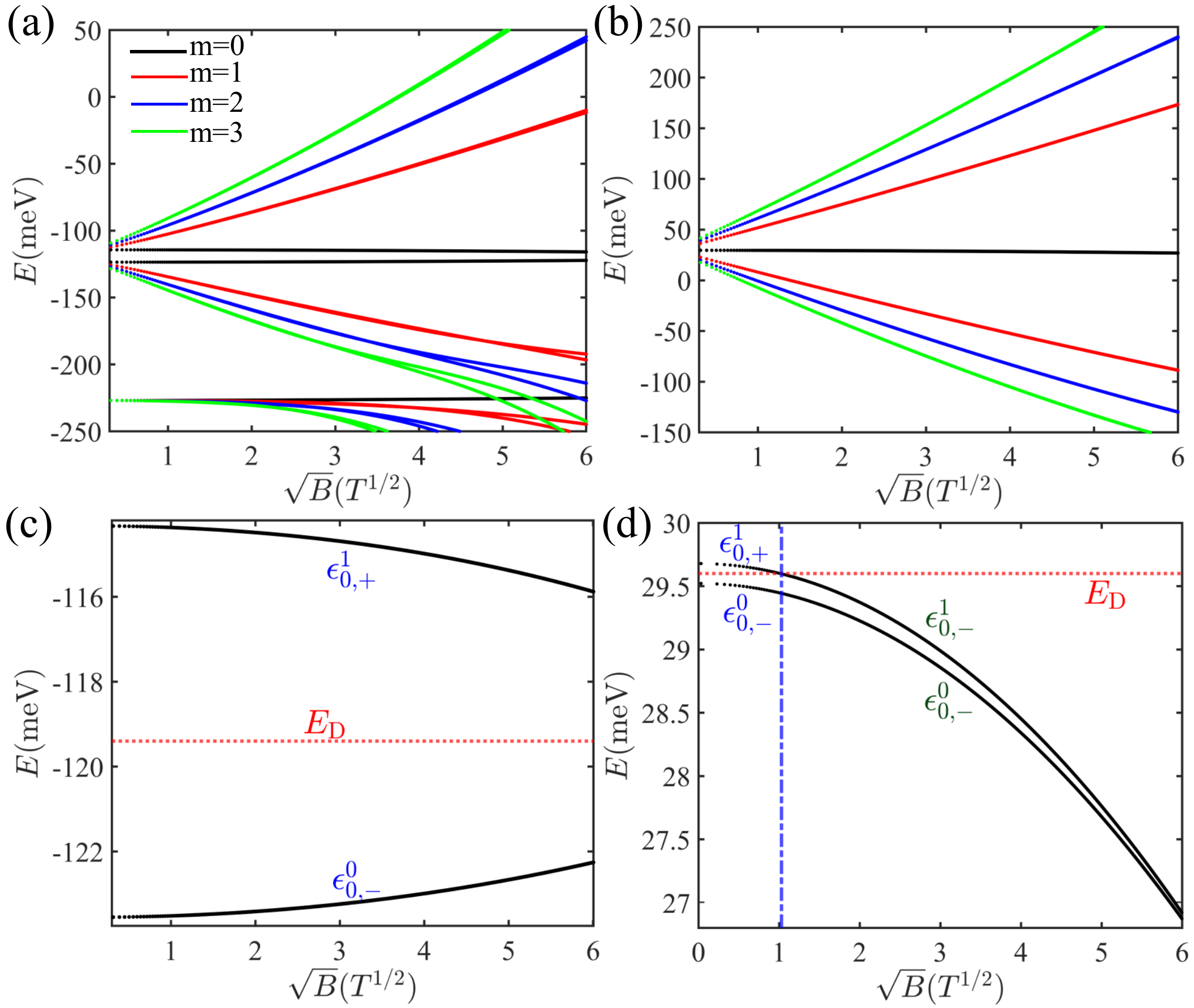}
  \caption{Landau level energy, $E$, as a function of $\sqrt{B}$ for an $8$QL thin-film of (a) Sb$_2$Te$_3$ and (b) Bi$_2$Se$_3$. Panels (c) and (d) show the dependence of the zeroth surface LL on $\sqrt{B}$ for the thin-films in (a) and (b).}\label{fig3}
\end{figure}
the magnetic length, and $y_{0}=k_{x}l^{2}_{\rm B}$ the guiding center coordinate. We  then follow the discretization procedure in Eq.~\eqref{discH}, and use the discrete spinor basis,
\begin{equation}\label{basisn}
|\overline{n}(z_{j})\rangle=\left[
                         \begin{array}{cc}
                           \hat{C}_{n,\uparrow,j}||n|-1\rangle & \hat{C}_{n,\downarrow,j}||n|\rangle  \\
                         \end{array}
                       \right]^{T}\;,
\end{equation}
where $n$ is the Landau level (LL) index, $n\ne0$, $\hat{C}_{n,\uparrow,j}= [a_{n,j}, b_{n,j}]$, $\hat{C}_{n,\downarrow,j}= [c_{n,j}, d_{n,j}]$ and $|n\rangle$ is the Fock state. This simplifies the boundary value problem to diagonalizing a block-tridiagonal matrix for each LL, $n$. For the LLs with $n \ne 0$ this matrix reads
\begin{eqnarray}\label{eigpn}
&&\left[H_{{\rm TF,B}}\right]^{(n)}_{l,j}=\\
&&\sum_{\alpha=\pm1}\sigma_{0}\left[-\tau_{0}\frac{D_{1}}{h^{2}_{z}}+\tau_{z}\frac{B_{1}}{h^{2}_{z}} -i\alpha\tau_{y}\frac{A_{1}}{2h_{z}}\right]\delta_{l+\alpha,j}\nonumber\\
&&+\left\{\mathcal{P}_{+}\left[\tau_{z}\mathcal{A}_{n}+\tau_{0}\mathcal{C}_{n}\right]+ \mathcal{P}_{-}\left[\tau_{z}\mathcal{B}_{n}+\tau_{0}\mathcal{D}_{n}\right]+\sigma_{y}\tau_{x}\mathcal{L}_{n}\right\}\delta_{l,j}\nonumber\;,
\end{eqnarray}
with $\mathcal{P}_{\pm}=(\sigma_{0}\pm \sigma_{z})/2$, $\mathcal{A}_{n}=M-B_{+}-\mu_{\rm B}Bg_{-}/4+2B_{1}/h^{2}_{z}$, $\mathcal{B}_{n}=M-B_{-}+\mu_{\rm B}Bg_{-}/4+2B_{1}/h^{2}_{z}$, $\mathcal{C}_{n}=C+D_{+}-\mu_{\rm B}Bg_{+}/4-2D_{1}/h^{2}_{z}$, $\mathcal{D}_{n}=C+D_{-}+\mu_{\rm B}Bg_{+}/4-2D_{1}/h^{2}_{z}$, $\mathcal{L}_{n}=A_{2}\sqrt{2|n|}/l_{\rm B}$, and we have defined $B_{\pm}=2B_{2}(|n|\pm1/2)/l_{\rm B}$ and $D_{\pm}=2D_{2}(|n|\pm1/2)/l_{\rm B}$. The eigenvalues of the matrix in Eq.~\eqref{eigpn} are the $\pm n$ LLs energies, $\epsilon_{\pm n}$, and their eigenvectors determine the coefficients $a_{\pm n}(z),\;b_{\pm n}(z),\;c_{\pm n}(z)$ and $d_{\pm n}(z)$. The $n=0$ LL is spin-polarized such that, $|\overline{0}(z_{j})\rangle= (0,0,c_{0,j}|0\rangle, d_{0,j}|0\rangle)$, and diagonalizing
\begin{eqnarray}\label{eigp0}
&&\left[H_{{\rm TF,B}}\right]^{(0)}_{l,j}=\left\{\left[\tau_{z}\mathcal{B}_{0}+\tau_{0}\mathcal{D}_{0}\right]+i\tau_{x}\mathcal{L}_{0}\right\}\delta_{l,j}\\
&&+\sum_{\alpha=\pm1}\left[-\tau_{0}\frac{D_{1}}{h^{2}_{z}}+\tau_{z}\frac{B_{1}}{h^{2}_{z}} -i\alpha\tau_{y}\frac{A_{1}}{2h_{z}}\right]\delta_{l+\alpha,j}\;.
\end{eqnarray}
determines $0^{\rm th}$ LL energy, $\epsilon_{0}$, and the coefficients $c_{0}(z)$ and $d_{0}(z)$. It is important to note that in order to capture all the properties of the LLs we assigned them two additional labels. First $s=\pm$, where $n=s m$ and $m>0$, describes the electron-like and hole-like nature of the LLs, where $s=+$ ($s=-$) if the LL energy is above (below) $E_{\rm D}$. Second, $\alpha=(0,1)$ classifies the LLs according to their $z$-inversion symmetry, {\it i.e.}, symmetric states ($\alpha=0$) and anti-symmetric states ($\alpha=1$), where $\mathcal{I}|\overline{m},s,\alpha,z\rangle$$=$$(-1)^{\alpha}|\overline{m},s,\alpha,z\rangle$=$|\overline{m},s,\alpha,-z\rangle$ and  $\mathcal{I}=\tau_{z}\sigma_{z}$ is the $z$-inversion operator. Hence the LLs of the TI thin-film system take the general spinor form
\begin{eqnarray}\label{LLsgen}
&&|\overline{m},s,\alpha,z\rangle=\\
&&\left[a^{\alpha}_{m,s}(z)|m-1\rangle\; b^{\alpha}_{m,s}(z)|m-1\rangle
                        \; c^{\alpha}_{m,s}(z)|m\rangle \; d^{\alpha}_{m,s}(z)|m\rangle\right]^{T},\nonumber
\end{eqnarray}
where $m$$>$$0$, and for the ${\rm 0}$$^{\rm th}$ LL $|\overline{0},s,\alpha,z\rangle$$=$$[0\; 0$$\; c^{\alpha}_{0,s}(z)|m\rangle \; d^{\alpha}_{0,s}(z)|m\rangle]^{T}$.

Diagonalizing Eqs.~\eqref{eigpn} and \eqref{eigp0} yields the surface and bulk LLs energies and states. In Fig.~\ref{fig3} we compare the LLs energies for an 8QL thin-film  of Sb$_2$Te$_3$, shown in panel (a), and Bi$_2$Se$_3$, in panel (b). For a TI thin-film the $m=0$ surface LL is split due to the presence of the Zeeman effect and the hybridization gap~\cite{burkov}. When $\Delta_{\rm H}(d)>\Delta_{\rm Z}(B)$, the surface LL energies for $m=0$ are located above and below the Dirac point energy, $E_{\rm D}=C+D_{1}M/B_{1}$, such that $\epsilon^{\alpha=1(0)}_{0,s=+(-)}=E_{\rm D}+s|\Delta_{\rm Z}(B)+(-1)^{\alpha} \Delta_{\rm H}(d)|$. In the latter expression $\Delta_{\rm Z}(B)=\mu_{\rm B}Bg_{+}/4$ is the Zeeman splitting, $\Delta_{\rm H}(d)=\Delta_{{\rm TF}}(d)/2$ where $\Delta_{{\rm TF}}(d)$ is given in Eq.~\eqref{agap}. The regime $\Delta_{\rm H}(d)>\Delta_{\rm Z}(B)$ is satisfied in the Sb$_2$Te$_3$ thin-film since $g_{1}\approx -g_{2}$, leading to $g_{+}\approx0$, and $\Delta_{\rm H}(d)>\Delta_{\rm Z}(B)$ for all values of $B$ considered in Fig.~\ref{fig3}. Hence, for an 8QL Sb$_2$Te$_3$ thin-film, the $m=0$ surface LL energies are $\epsilon^{\alpha}_{0,s}\approx E_{\rm D}+s \Delta_{\rm H}(d)$, as shown in Fig.~\ref{fig3}(c). For an 8QL Bi$_2$Se$_3$ thin-film, at small values of the magnetic field we have $\Delta_{\rm H}(d)>\Delta_{\rm Z}(B)$, and the zeroth surface LL energies, $\epsilon^{\alpha}_{0,s}$, are similar to their Sb$_2$Te$_3$ counterpart.
However $\Delta_{\rm Z}(B)$ rapidly becomes larger than $\Delta_{\rm H}(d)$ since the hybridization gap for the Bi$_2$Se$_3$ thin-film is almost negligible, Fig.~\ref{fig3}(c), and $g_{+}>1$. For $B	\gtrsim 1$T the energies of the $m=0$ surface LL become located below the Dirac point and correspond to $\epsilon^{\alpha}_{0,s=-}=E_{\rm D}+s|\Delta_{\rm Z}(B)+(-1)^{\alpha} \Delta_{\rm H}(d)|$. Another interesting difference between the Bi$_2$Se$_3$ and Sb$_2$Te$_3$ thin-film arises from the strong p-h asymmetry displayed in Sb$_2$Te$_3$ [see Fig.~\ref{fig1}(b) and Fig.~\ref{fig2}(b)]. The p-h symmetry breaking in Sb$_2$Te$_3$  makes the gap center ($E_{\rm D}$ in the gapless case) close to the bulk-valence band, Fig.~\ref{fig2}(b), leading to bulk LLs closer in energy to the surface states LLs, as shown in Fig.~\ref{fig3}(a). However, in Bi$_2$Se$_3$ the Dirac point is energetically distant from both the bulk-valence and bulk-conduction bands, Fig.~\ref{fig2}(c). Hence, for the Bi$_2$Se$_3$ thin-film we notice that for the rather large window of energies shown in Fig.~\ref{fig3}(c) only the surface LLs appear due to the relatively large energy difference between them and the bulk LLs. In what follows we consider the magneto-optical response of TI thin-films and investigate the effects of p-h asymmetry of the surface and bulk LLs on the dynamical conductivity.

\section{Magneto-optical Conductivity}\label{sec3}

\begin{figure}[t]
  \centering
  \includegraphics[width=\columnwidth]{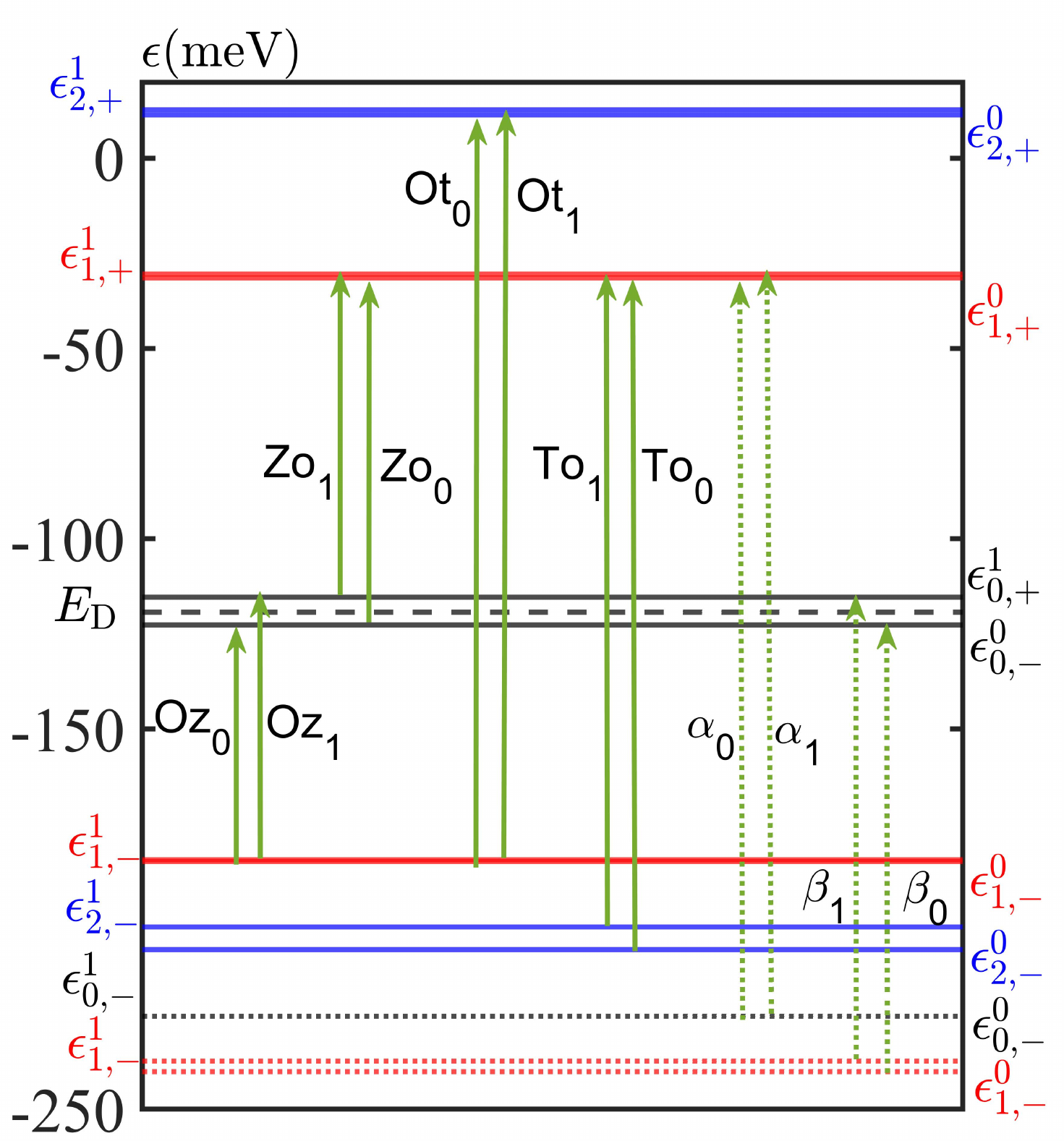}
  \caption{Landau levels of an $8$QL Sb$_2$Te$_3$ thin-film subjected to a $B=25$T magnetic field. The solid horizontal lines represent the surface LLs, the dotted horizontal lines are the bulk LLs, and the solid (dotted) arrows indicate the non-vanishing surface-to-surface (bulk-to-surface) optical transitions. Our labelling notation of the optical transitions is explained in~\cite{note2}.}\label{fig4}
\end{figure}

In this section we evaluate the magneto-optical conductivity of a TI thin-film. Adopting the self-consistent method to find the bulk and surface LLs, Eqs.~\eqref{basisn}-\eqref{eigpn}, we identify the effects of quantum confinement and p-h symmetry breaking on the magneto-optical conductivity of the system. In the Landau level representation, the Kubo formula for the optical conductivity tensor, $\sigma_{ab}(\omega)$, is given by ~\cite{WKT}
\begin{eqnarray}\label{cond}
&&\sigma_{ab}(\omega)=\frac{i\hbar}{2\pi l_{B}^{2}}\sum_{m,m'}\sum_{\substack{s,s'=\pm\\\alpha, \alpha'=(0,1)}} \frac{f^{\alpha}_{m,s}-f^{\alpha'}_{m',s'}}{\left(\epsilon^{\alpha}_{m,s}-\epsilon^{\alpha'}_{m',s'}\right)}\bigg[\\
&&\frac{\langle \alpha,s,\overline{m}| J_{a}|\overline{m'},s',\alpha '\rangle \langle \alpha ',s',\overline{m}'| J_{b}|\overline{m},s,\alpha\rangle}{\left(\omega+\epsilon^{\alpha}_{m,s}-\epsilon^{\alpha'}_{m',s'}+i\eta\right)}\bigg]\;.\nonumber
\end{eqnarray}
Here
\begin{equation}\label{matrixelementint}
\langle \alpha,s,\overline{m}| J_{a}|\overline{m'},s',\alpha '\rangle =\int_{-a}^{a}dz\langle z, \alpha,s,\overline{m}| J_{a}|\overline{m'},s',\alpha ', z\rangle\;,
\end{equation}
$d=2a$ is the thin-film thickness, $f^{\alpha}_{m,s}=\Theta(E_{\rm F}-\epsilon^{\alpha}_{m,s})$ is the zero-temperature Fermi-Dirac distribution function, and $E_{\rm F}$ is the Fermi level. Moreover, $\eta$ is the TI's quasiparticles lifetime broadening taken to be small compared to typical Landau level spacings in the frequency range considered.
The current operator ${\bm J}=(J_{x},J_{y})$ is
\begin{eqnarray}\label{currentops}
  && J_{x} =\frac{2 e}{\hbar}\sigma_{0}\left(\tau_{z}B_{2}-\tau_{0}D_{2}\right)\left(\frac{a+a^{\dag}}{\sqrt{2}l_{B}}\right)+\frac{e A_{2}}{\hbar}\sigma_{x}\tau_{x}\nonumber \\
  && J_{y} = \frac{2i e}{\hbar}\sigma_{0}\left(\tau_{z}B_{2}-\tau_{0}D_{2}\right)\left(\frac{a-a^{\dag}}{\sqrt{2}l_{B}}\right)+\frac{e A_{2}}{\hbar}\sigma_{y}\tau_{x}\;.\\
  &&\nonumber
\end{eqnarray}
The current operator matrix elements $\langle \alpha,s,\overline{m}| J_{a}|\overline{m'},s',\alpha '\rangle$ in Eq.~\eqref{matrixelementint} satisfy the selection rule $|sm|-|s'm'|=\pm1$ with $\alpha=\alpha'$. In Fig.~\ref{fig4} we numerically verify the selection rule by considering an 8QL Sb$_2$Te$_3$ thin-film exposed to a $25$T magnetic field.

\begin{figure*}[t]
  \centering
  \includegraphics[width=\textwidth]{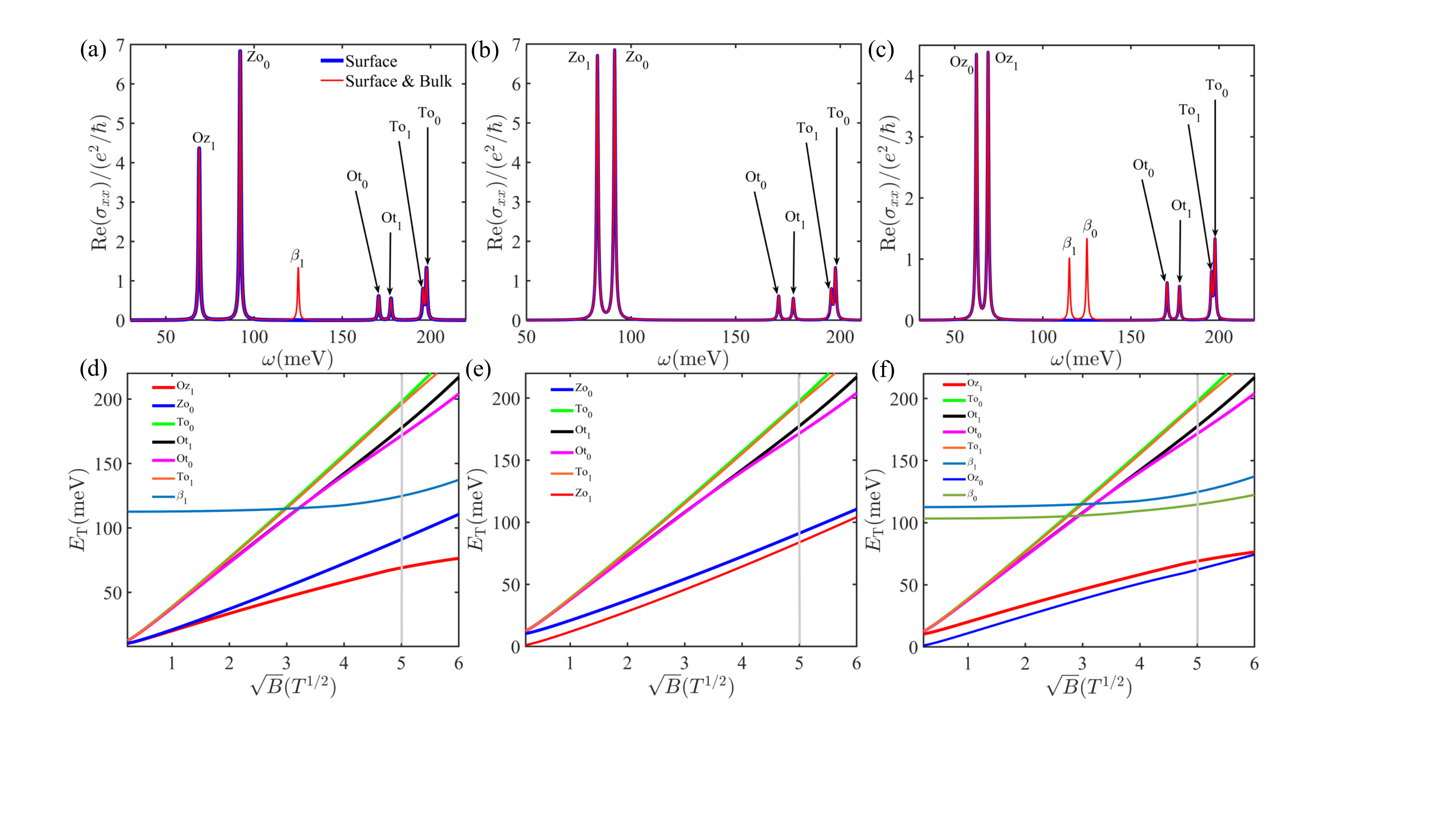}
  \caption{Real part of the magneto-optical conductivity of an $8$QL Sb$_2$Te$_3$ thin-film subjected to a $B=25$T field for (a) undoped sample, $E_{\rm F}=E_{\rm D}$, (b) n-doped, $E_{\rm F}=E_{\rm D}+\delta_{\rm E}$ and (c) p-doped sample with $E_{\rm F}=E_{\rm D}-\delta_{\rm E}$ ($E_{\rm F}$ is the Fermi level, $E_{\rm D}$ is the Dirac point, and $\delta_{\rm E}=50$ meVs). Panels (d)-(f) show the dependence of the optical transition energy on $\sqrt{B}$ for the cases in (a)-(c).  }\label{fig5}
\end{figure*}

Numerically evaluating Eq.~\eqref{cond} we find the absorptive part of the longitudinal conductivity, ${\rm Re}[\sigma_{xx}(\omega)]$. For an 8QL thin-film of Sb$_2$Te$_3$ at 25 T we find  ${\rm Re}[\sigma_{xx}(\omega)]$ for three different doping levels, Fig.~\ref{fig5}(a)-(c). In our calculation of ${\rm Re}[\sigma_{xx}(\omega)]$, we first include only the contributions of the surface states and then we compare with a full calculation including both surface and bulk states. This procedure allows the identification of bulk states contributions to ${\rm Re}[\sigma_{xx}(\omega)]$. The non-vanishing LL transitions in the Sb$_2$Te$_3$ thin-film are shown in Fig.~\ref{fig4} and we label each of these transitions for concreteness. In Ref.~\cite{note2} we describe the notation of our labeling system of the energy transitions. When the Fermi level is located at the Dirac point, Fig.~\ref{fig5}(a), we notice that the lowest transition energy ($|\epsilon^{\alpha}_{m,s}-\epsilon^{\alpha'}_{m',s'}|$), Fig.~\ref{fig4}, is that of ${\rm Oz_{1}}$, {\it i.e.}, between the first ($m=1$) hole-like ($s=-$) asymmetric ($\alpha=1$) surface LL and the zeroth, electron-like, asymmetric surface LL. The second low energy transition allowed is between the surface LLs $(m=0,s=-,\alpha=0)$ and $(m=1,s=+,\alpha=0)$, ${\rm Zo_{0}}$ in Fig.~\ref{fig4}. Surprisingly, the energy of the bulk-to-surface transition $\beta_{1}$ is lower than all other allowed surface-to-surface optical transitions involving LLs with $m>0$. This bulk-to-surface transition,
\begin{figure*}[t]
  \centering
  \includegraphics[width=\textwidth]{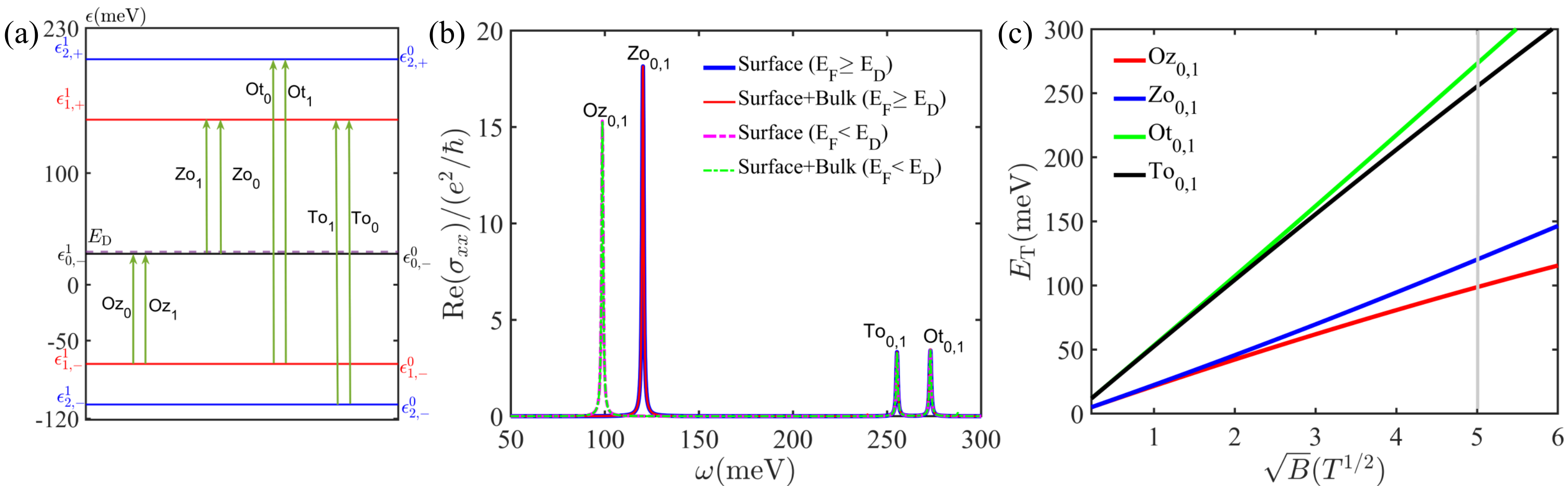}
  \caption{Panel (a) shows the LLs of an 8QL Bi$_2$Se$_3$ thin-film subjected to a $25$T magnetic field. The arrows in this panel indicate the non-vanishing optical transitions. The conventions of this panel are same as in Fig.~\ref{fig4}. (b) Absorption part of the magneto-optical conductivity of the thin-film in (a) for three different Fermi level locations with respect to the Dirac point (same as Fig.~\ref{fig5}). (c) Dependence of the optical transition energies on $\sqrt{B}$. The transitions allowed for the three different doping levels considered in panel (b) are displayed.}\label{fig6}
\end{figure*}
$\beta_{1}$, happens between the $(m=1,s=-,\alpha=1)$ bulk LL and the $(m=0,s=+,\alpha=1)$ surface LL. From low to high energy the remaining surface-to-surface transitions in Fig.~\ref{fig4} are ${\rm Ot_{0}}$, ${\rm Ot_{1}}$, ${\rm To_{1}}$ and ${\rm To_{0}}$. We also note that the energy  of bulk-to-surface transitions $\alpha_{1,2}$ are higher than all surface-to-surface transition energies shown in Fig.~\ref{fig4}. The absorptive part of the longitudinal conductivity, ${\rm Re}[\sigma_{xx}(\omega)]$, for $E_{\rm F}=E_{\rm D}$ displays resonant peaks at the location of the allowed transitions for this doping level, as shown in Fig.~\ref{fig5}(a). When the Fermi level migrates above the Dirac point, $50$ meVs above $E_{\rm D}$ in Fig.~\ref{fig5}(b), we notice the disappearance of the peaks corresponding to the ${\rm Oz_{1}}$ surface-to-surface and $\beta_{1}$ bulk-to-surface transitions in ${\rm Re}[\sigma_{xx}(\omega)]$, since the $(m=0,s=+,\alpha=1)$ surface LL becomes occupied at this doping level (see Fig.~\ref{fig4}). On the other hand, ${\rm Re}[\sigma_{xx}(\omega)]$ displays a peak at the ${\rm Zo_{1}}$ transition energy due to the occupation of both $m=0$ LLs, see Fig.~\ref{fig4}. Moreover, in the doping regime of Fig.~\ref{fig5}(b) ($E_{\rm F}>E_{\rm D}$), all the resonant peaks of the longitudinal conductivity are associated with surface-to-surface transitions for the relevant range of incident photons energy, $\omega$. When the Fermi level is below the Dirac point, Fig.~\ref{fig5}(c) where $E_{\rm F}$ is 50 meV below $E_{\rm D}$, the lowest energy resonant peaks locations in ${\rm Re}[\sigma_{xx}(\omega)]$ become those associated with ${\rm Oz_{0}}$ and ${\rm Oz_{1}}$ transitions. The latter is due to the depopulation of the $m=0$ surface LLs which also leads to the absence of the peaks at ${\rm Zo_{0,1}}$ (see Fig.~\ref{fig4}). In this case the absorption peaks of $\sigma_{xx}(\omega)$ associated with surface-to-bulk transitions increase and appear at even lower photon energies when compared to the case in Fig.~\ref{fig5}(a). The new surface-to-bulk peak in ${\rm Re}[\sigma_{xx}(\omega)]$ is associated with $\beta_{0}$, Fig.~\ref{fig4}, which becomes allowed since for $E_{\rm F}<E_{\rm D}$  the $(m=0,s=-1,\alpha=0)$ surface LL is not occupied. The remaining absorption peaks for $E_{\rm F}<E_{\rm D}$ are similar to those in the previous two cases in Figs.~\ref{fig5}(a) and (b). The low photon energies at which the bulk-to-surface peaks are displayed in the undoped and $n$-doped cases, in Figs.~\ref{fig5}(a) and (c), are attributed to the strong p-h asymmetry of Sb$_2$Te$_3$. The latter brings the surface LLs in close proximity to the bulk-valence LLs, as shown in Fig.~\ref{fig3}(a). Interestingly, TI thin-films doping leads to an asymmetric enhancement of the bulk contributions in the optical conductivity, as seen in Figs.~\ref{fig5}(a)-(c), and provides a direct indication of the asymmetry between electrons and holes in the system.

Experimentally the transition energy dependence on the magnetic field is extracted from the  magneto-optical (usually infrared) spectrum~\cite{yuxuan,dcond3,dcond}. In Figs.~\ref{fig5}(d)-(f) we show the transition energies corresponding to the doping levels in Fig.~\ref{fig5}(a)-(c). In the transition energy plots in Figs.~\ref{fig5}(d)-(f), we notice that the surface-to-surface transition displays a linear dependence with the $\sqrt{B}$, {\it i.e.}, $E_{\rm T}\approx \hbar v_{\rm F}\sqrt{2e B/\hbar} |s\sqrt{m+1}+s'\sqrt{m}|$, where $v_{\rm F}$ is the Fermi velocity of the surface states. The latter is typical for Dirac fermions and the relatively small deviations of these transitions away from their linear dependence on $\sqrt{B}$ is ascribed to the quadratic dependence enhanced by p-h asymmetry~\cite{magopt2}. In the undoped and the n-doped cases, Figs.~\ref{fig5}(a) and (c), the bulk-to-surface optical transitions are easily identifiable due to their peculiar dependence on $\sqrt{B}$ which does not follow the typical linear dependence of the surface-to-surface optical transitions. Additionally, the low energy character of the bulk-to-surface transition and their enhancement with p-doping indicates the proximity of the surface state's Dirac point to the bulk-valence band due to the strong p-h asymmetry in this material. Two important conclusions about the absorptive part of the longitudinal conductivity of Sb$_2$Te$_3$ can be drawn from the optical transitions energy dependence on $\sqrt{B}$; first, when $E_{\rm F}\le E_{\rm D}$ the bulk-to-surface contributions will be displayed at low photon frequencies even for small values of $B$. Second, since the contribution of the bulk-to-surface transitions arise from the strong p-h asymmetry that shifted the Dirac point of the surface states close to the bulk-valence, n-doping of Sb$_2$Te$_3$ will effectively counteract the effects of p-h asymmetry and suppress the bulk-to-surface contribution in ${\rm Re}[\sigma_{xx}(\omega)]$, Fig.~\ref{fig5}(b) and (e).

In Bi$_2$Se$_3$ the p-h asymmetry is less pronounced as its Dirac point remains relatively far from both the conduction and valence-bulk bands, Fig.~\ref{fig2}. Additionally, since the hybridization gap of an 8QL  Bi$_2$Se$_3$ thin-film is smaller than the Zeeman interaction energy for a 25T magnetic field, the symmetric and antisymmetric $0^{\rm th}$ surface LLs have a hole-like character and are approximately degenerate, as shown in Fig.~\ref{fig3}(d). Similarly, for higher LLs, $m=1,2\dots$, the difference in energy between their symmetric and antisymmetric states is also negligible, Fig.~\ref{fig6}(a). The absorptive part of $\sigma_{xx}(\omega)$ is shown in Fig.~\ref{fig6}(b) for three different locations of the Fermi level. For the cases where $E_{\rm F}=E_{\rm D}$ and $E_{\rm F}$ is above the Dirac point, we notice identical absorption peaks in ${\rm Re}[\sigma_{xx}(\omega)]$. This can be understood from the LLs and transition energies in Fig.~\ref{fig6}(a). Since the $m=0$ symmetric and antisymmetric surface LLs are below the Dirac point then the optical transitions ${\rm Zo_{0,1}}$ are the lowest energy transitions in this case. Shifting $E_{\rm F}$ above the Dirac point (50meVs above $E_{\rm D}$ in Fig.~\ref{fig6}) does not introduce new optical transitions since both $m=0$ surface LLs remain populated while the surface LLs at $\epsilon^{0,1}_{1,+}$ remain unpopulated. Conversely, when the Fermi level is shifted below $E_{\rm D}$ (50meVs below $E_{\rm D}$ in Fig.~\ref{fig6}) the $m=0$ surface LLs are emptied. The shift of $E_{\rm F}$ below the $m=0$ surface LLs makes ${\rm Oz_{0,1}}$ the lowest energy transition while removing the ${\rm Zo_{0,1}}$ transition. The remaining transitions ${\rm Ot_{0,1}}$ and ${\rm To_{0,1}}$ appear for all the doping levels considered. The interplay between the doping level and the transition energies, Fig.~\ref{fig6}(a), is reflected in the photon energies leading to absorption peaks in ${\rm Re}[\sigma_{xx}(\omega)]$, Fig.~\ref{fig6}(b). The dependence of the transition energies on the $\sqrt{B}$ in Fig.~\ref{fig6}(c) show the typical linear dependence, $E_{\rm T}=\hbar v_{\rm F}\sqrt{2e B/\hbar} |s\sqrt{m+1}+s'\sqrt{m}|$, of massless Dirac fermions. The latter reflects the surface-to-surface nature of the optical transitions. Moreover, for  Bi$_2$Se$_3$ the small shift in the Dirac point resulting from p-h symmetry breaking leads to the absence of the bulk-to-surface contributions in the longitudinal optical conductivity for a rather large range of light frequencies and magnetic fields, as shown in Figs.~\ref{fig6}(b) and (c).

In this section, we have demonstrated that self-consistently obtaining the surface and bulk LLs arising in a TI thin-film subjected to a quantizing magnetic field, and considering  realistic material parameters that capture intrinsically broken symmetries are essential to understanding the magneto-optical response of these TI nanostructures~\cite{yuxuan}. We have shown that the strong contrast between the dynamic longitudinal conductivity of Sb$_2$Te$_3$ and Bi$_2$Se$_3$ arises from the different levels of p-h asymmetry in these materials. While in Sb$_2$Te$_3$ p-h asymmetry leads to a considerable shift of the Dirac point towards the bulk-valence band, in Bi$_2$Se$_3$ the Dirac point remains roughly equidistant from the bulk-valence band and the bulk conduction band. Hence, when subjected to a perpendicular magnetic field and an infrared light, Sb$_2$Te$_3$ will display bulk contributions in its magneto-optical conductivity, and these contributions are enhanced with p-doping and become suppressed with n-doping since it counteracts p-h asymmetry. In Bi$_2$Se$_3$ the absorption component of the longitudinal magneto-optical conductivity arises solely from the surface LLs for a relatively large range of photon energies.

\section{Summary and conclusions }\label{sec4}
We have studied the effects of particle-hole (p-h) symmetry breaking and quantum confinement on the magneto-optical response of topological thin-films. We showed that similar to the surface states and their extension into the TI bulk, the hybridization gap generated in TI thin-films strongly depends on the intrinsic p-h asymmetry. The effects of quantum confinement are taken into account by a self-consistent approach to the thin-film problem, {\it i.e.}, simultaneously considering the states that satisfy the TI bulk Hamiltonian and the termination imposed boundary condition. Our analysis of the TIs Sb$_2$Te$_3$ and Bi$_2$Se$_3$ showed that Sb$_2$Te$_3$ displays a considerable p-h asymmetry since the Dirac point of its surface states is pushed closer to the valence-bulk band, while in Bi$_2$Se$_3$ the Dirac point remains relatively far from the bulk conduction and valence bands. In the presence of a quantizing magnetic field, we find the surface and bulk Landau levels (LLs) of a TI thin-film. We show that p-h symmetry breaking is also reflected in the LLs spectrum. For Sb$_2$Te$_3$ thin-films the surface states LLs are energetically close to the bulk-valence LLs while for Bi$_2$Se$_3$, the surface LLs remain energetically distant from the bulk LLs.

Our analysis of the absorptive part of the longitudinal conductivity provided a clear contrast between TI thin-films with different levels of p-h asymmetry. The sizable breaking of p-h symmetry in TI thin-films of Sb$_2$Te$_3$ was reflected by low photon energy absorption peaks arising from bulk-to-surface LLs transitions in the longitudinal optical conductivity. In thin-films with a small degree of p-h symmetry breaking, Bi$_2$Se$_3$, the bulk-to-surface LLs optical transitions are absent for a sizable range of photon-energies. Additionally, we show that the asymmetric enhancement of bulk-to-surface transitions and their associated absorption peaks unveil the asymmetric location of the Dirac point. For example, for a thin-film of Sb$_2$Te$_3$ p-doping enhances the absorption peaks from bulk-to-surface LLs transitions, while n-doping reduces these peaks in the longitudinal conductivity. The latter happens because the p-h asymmetry in Sb$_2$Te$_3$ pushed the Dirac point close to the valence band. On the other hand, in Bi$_2$Se$_3$ thin-films, n-doping and p-doping did not lead to a skewed enhancement of the absorption peaks arising from bulk-to-surface LLs transitions in the dynamic conductivity, which indicated that the Dirac point in this material is, approximately, symmetrically located with respect to the bulk conduction and valence bands. In addition to the indications provided by the absorption peaks in the longitudinal conductivity, we showed that the dependence of the transition energies on the strength of the magnetic field provides clear indications of the appearance of bulk contributions to the LLs transitions. The bulk-to-surface LLs transitions are easily distinguished from surface-to-surface transitions due to their non-linear dependence on the square-root of the magnetic field amplitude. Our theory and observations in this work provide a clear guide to experimentally discerning particle-hole asymmetry in TI thin-films via magneto-optical measurements ~\cite{yuxuan}.

\acknowledgments{
We gratefully acknowledge Yuxuan Jiang and Liang Wu for fruitful discussions leading to this work. This work was supported by the U.S. Department of Energy, Office of Science, Basic Energy Sciences under Early Career Award No.~DE-SC0019326.}

\bibliographystyle{apsrev4-2}
\bibliography{refsTI}
\end{document}